\documentclass[lettersize,journal]{IEEEtran}
\usepackage{amsmath,amsfonts}
\usepackage{algorithmic}
\usepackage{array}
\usepackage{cite}
\usepackage[caption=false,font=normalsize,labelfont=sf,textfont=sf]{subfig}
\usepackage{textcomp}
\usepackage{stfloats}
\usepackage{multirow}
\usepackage{url}
\usepackage{verbatim}
\usepackage{graphicx}
\usepackage{xcolor}
\usepackage[switch]{lineno}
\def\BibTeX{{\rm B\kern-.05em{\sc i\kern-.025em b}\kern-.08em
    T\kern-.1667em\lower.7ex\hbox{E}\kern-.125emX}}
\usepackage{balance}
\begin{document}
\title{A Generative Shape Compositional Framework to Synthesise Populations of Virtual Chimaeras}
\author{Haoran Dou, Seppo Virtanen, Nishant Ravikumar$^*$, Alejandro F. Frangi$^*$, \IEEEmembership{Fellow, IEEE}
\thanks{
This work was supported in part by the Royal Academy of Engineering INSILEX Chair (CiET1919/19), UKRI Frontier Research Guarantee INSILICO (EP\textbackslash Y030494\textbackslash 1), and EPSRC (EP/W007819/1). 

Haoran Dou and Nishant Ravikumar are with the Center for Computational Imaging and Simulation Technologies in Biomedicine within the School of Computing at the University of Leeds, LS2 9JT Leeds, UK. (schd@leeds.ac.uk; n.ravikumar@leeds.ac.uk)

Seppo Virtanen is with the Center for Computational Imaging and Simulation Technologies in Biomedicine within the School of Mathematics at the University of Leeds, LS2 9JT Leeds, UK. (s.virtanen@leeds.ac.uk)

Alejandro F. Frangi is with the Christabel Pankhurst Institute, Division
of Informatics, Imaging and Data Sciences, University of Manchester,
M1 3BB Manchester, UK, and also with the Department of Computer
Science, University of Manchester, M1 3BB Manchester, U.K., and also with the Medical Imaging Research Center (MIRC), Departments of Electrical Engineering and Cardiovascular Sciences, KU Leuven, Leuven, Belgium, and also with Alan Turing Institute, London, UK (alejandro.frangi@manchester.ac.uk). 

* Nishant Ravikumar and Alejandro F. Frangi are joint last authors.

Corresponding author: Alejandro F. Frangi.

© 2024 IEEE.  Personal use of this material is permitted.  Permission from IEEE must be obtained for all other uses, in any current or future media, including reprinting/republishing this material for advertising or promotional purposes, creating new collective works, for resale or redistribution to servers or lists, or reuse of any copyrighted component of this work in other works.
}}

\markboth{IEEE Trans. Neural Netw. Learn. Syst., 2024, in press}%
{How to Use the IEEEtran \LaTeX \ Templates}

\maketitle

\begin{abstract}
Generating virtual organ populations that capture sufficient variability while remaining plausible is essential to conduct \textit{in-silico} trials of medical devices. However, not all anatomical shapes of interest are always available for each individual in a population. The imaging examinations and modalities used can vary between subjects depending on their individualised clinical pathways. Different imaging modalities may have various fields of view, are sensitive to signals from other tissues/organs, or both. Hence, missing/partially overlapping anatomical information is often available across individuals. We introduce a generative shape model for multipart anatomical structures, learnable from sets of unpaired datasets, i.e., where each substructure in the shape assembly comes from datasets with missing or partially overlapping substructures from disjoint subjects of the same population. The proposed generative model can synthesise complete multipart shape assemblies coined \textit{virtual chimaeras}. We applied this framework to build \textit{virtual chimaeras} from databases of whole heart shape assemblies that each contribute samples for heart substructures. Specifically, we propose a graph neural network-based generative shape compositional framework which comprises two components, a part-aware generative shape model which captures the variability in shape observed for each structure of interest in the training population, and a spatial composition network which assembles/composes the structures synthesised by the former into multipart shape assemblies (viz. \textit{virtual chimaeras}). We also propose a novel self-supervised learning scheme that enables the spatial composition network to be trained with partially overlapping data and weak labels. We trained and validated our approach using shapes of cardiac structures derived from cardiac magnetic resonance images in the UK Biobank. When trained with \textit{complete} and \textit{partially overlapping} data, our approach significantly outperforms a PCA-based shape model (trained with \textit{complete} data) in terms of generalisability and specificity. This demonstrates the superiority of the proposed method, as the synthesised cardiac virtual populations are more plausible and capture a greater degree of shape variability than those generated by the PCA-based shape model. 
\end{abstract}

\begin{IEEEkeywords}
In-silico trials, virtual populations, graph neural network, generative model.
\end{IEEEkeywords}

\section{Introduction}
\IEEEPARstart{I}{n}-\textit{silico} trials (IST) offer a paradigm shift for the innovation of medical devices and the regulatory approval process that underpins the path to the commercialisation of devices and their adoption in routine patient care. Traditionally, collecting scientific evidence for the regulatory approval of new devices and drugs required in vitro and in vivo evaluation of safety and efficacy. However, this practice poses a methodological and economic burden on medical products that impede innovation and delay/hinder patient benefit. In response, regulatory agencies are increasingly embracing complementary sources of evidence that refine, replace, and reduce the need for animal and human testing \cite{viceconti2021silico}. ISTs use computational modelling and simulation to assess the safety and efficacy of medical devices in the virtual world, in populations of digital twins or virtual patients (VP), and provide digital (or \textit{in silico}) as opposed to real-world evidence that supports device regulatory approval. Therefore, ISTs have the potential to explore device performance in a wider range of patient characteristics than it is feasible to recruit for in a real clinical trial. They could help refine, reduce, and partially replace \textit{in vivo} clinical trials.

ISTs require the generation of digital twin or virtual patient populations that capture sufficient anatomical and physiological variability, representative of the target patient populations, to allow a meaningful in silico evaluation of device performance~\cite{cheng2022virtual}. In this study, we focus on the challenge of generating representative populations of anatomy, specifically cardiovascular anatomy. We also define digital twins~\cite{tao2022digital} as distinct from virtual patients in the following way: digital twins are considered to be patient-specific replicas of anatomical structures generated by segmenting medical images (e.g. magnetic resonance (MR) or computed tomography (CT) images) of patients, and representing the shape of anatomical structures of interest in some parametric form (e.g. triangular surface meshes). Virtual patients~\cite{sinisi2020complete} are considered parametric representations of anatomical structures sampled from a generative model (e.g., probabilistic PCA, variational autoencoder (VAE)), where the latter was learnt from a population of digital twins. Virtual patients do not represent any specific patient's anatomy, but are instances generated from a trained model. We introduce \textit{virtual chimaeras} (VCs), as opposed to natural human chimaeras \cite{madan2020natural}, as distinct entities from VPs in the following way: a VC is a parametric representation of anatomical structures sampled from a generative model trained using \textit{partially overlapping} data, i.e. where all anatomical structures of interest are not available for all individuals in the training population~\cite{mitra2023learning, shi2021marginal, zhang2021dodnet}. This is distinct from VPs as we define them as instances of statistical/generative models that require \textit{complete overlap} in the training data, i.e. all anatomical structures of interest are available for all individuals in the training population (e.g., PCA). Henceforth, we will refer to these scenarios that lead to VCs and VPs as learning statistical/generative shape models with \textit{'partial overlap'} or \textit{'complete overlap'} in training data for brevity.

\subsection{Virtual Population Modelling}
Virtual populations of anatomical shapes (typically represented as computational meshes) are essential for conducting ISTs of medical devices. However, building rich/descriptive generative shape models of multipart anatomical structures is challenging. Most techniques require large volumes of training data that comprise the same semantic parts/shapes in each training sample~\cite{li2020learning,li2021editvae}. This would require expensive and laborious annotation of medical imaging data to ensure that anatomical shapes of interest can be accurately extracted from each subject/sample, which is prohibitive. Relying on complete annotations precludes using existing public databases where annotations for a few anatomical structures of interest may already be available~\cite{shi2021marginal,zhang2021dodnet}. Additionally, specific anatomical structures may only be visible/easy to delineate in specific image modalities that are less prevalent. For example, cine-MR images do not capture the 3D structure of the atria or the aorta within the field of view. However, large-scale databases such as the UK Biobank are available (comprising $>$40k subjects' images)~\cite{xia2022}. In contrast, computed tomography angiography (CTA) captures fine details of all four cardiac chambers and the associated great vessels (e.g. aorta and pulmonary artery) in 3D owing to its high spatial resolution. Still, it brings the added risk of patient exposure to radiation. Consequently, the scale of publicly available CTA data is usually much smaller than cine-MRI.

Although some studies in the computer vision domain~\cite{niemeyer2021giraffe,cao2018dida} have investigated learning multipart generative shape models using disparate data sets with non-/partially overlapped parts (i.e., where individual components from different data sets are leveraged to learn a model that synthesises multipart shape assemblies), this remains unexplored within the medical imaging domain to the best of our knowledge. Given the deluge of annotated medical imaging data (characterising anatomical structures of interest) that have been curated in recent years, which is likely to increase in coming years, there is a need in the medical image computing community to develop techniques that facilitate the estimation of multipart generative shape models, by using anatomical structures that may be available in multiple disparate populations/databases and image modalities. 

\subsection{Relevant Literature}
This study focusses on the generation of virtual populations of cardiovascular anatomy represented as triangular meshes of the surface. While several previous studies have proposed image-based generative models that capture anatomical variability across a people, we restrict our discussion of relevant literature to previous work on statistical/generative shape modelling. Early work on statistical shape models (SSMs) was motivated by the need for model-constrained image segmentation approaches to preserve the topology of segmented anatomical structures by using SSMs as a shape prior. However, a detailed review of these methods is beyond the scope of this study, and we refer the reader to the review by Heimann \textit{et al.}~\cite{heimann2009statistical} for more information on the topic.

PCA-based statistical shape modelling has thus far been the most common approach to building virtual populations of anatomical shapes. SSMs learnt using PCA were popularised by~\cite{cootes1995active} and have been used extensively to generate virtual anatomy populations~\cite{ravikumar2018, gooya2015bayesian}, in quantitative shape analysis for computer-aided diagnosis~\cite{shen2012detecting} and model-based segmentation~\cite{castro2015statistical} approaches. SSMs have also been extensively used in the cardiovascular domain to explore associations between cardiac morphology and function~\cite{young2009computational, rodero2021linking}, to segment cardiac images~\cite{pereanez2015patient, alba2015algorithm, attar2019quantitative}, and to characterise shape variability in healthy~\cite{bernardino2020handling} and pathological populations~\cite{lamata2013computational}. Inspired by~\cite{frangi2002automatic}, the first study to construct a multipart SSM of the heart, several subsequent studies~\cite{ordas2007computational, faghih20134d, perperidis2005construction} have used PCA and its variants to build 4D (3D+time) statistical models that can capture spatiotemporal variability between and within subjects in cardiac shape, simultaneously. For example, Hoogendoorn~\textit{et al.}~\cite{hoogendoorn2009bilinear} decoupled the shape between subjects and the temporal variations (dynamics) between subjects through a bilinear model, allowing extrapolation of the cardiac phase from the SSM even in the absence of individual measurements. 

PCA-based SSMs generate virtual shape populations by sampling from the shape space spanned by an orthogonal set of basis vectors, i.e. the eigenvectors (or principal components) of the covariance matrix of the population of shapes used to build the model. Both point-set- and mesh-based representations of anatomical structures have been used to create SSMs in this way. However, a prerequisite for PCA-based SSMs is point-wise correspondence across the population of training shapes (typically achieved via co-registration of shapes before SSM construction) and \textit{complete overlap} across all samples in the training population. In other words, standard PCA-based SSMs cannot handle missing structures or effectively use training samples with \textit{partial overlap} in anatomical structures. Recent advances in deep learning have shown that deep neural networks can be formulated as powerful generative models for images and geometric (e.g., point clouds, meshes/graphs) data due to their ability to learn rich hierarchical representations of data~\cite{harshvardhan2020comprehensive}. It has been widely adopted in different applications, such as 3D face reconstruction~\cite{ranjan2018generating, gao2022robust}, molecular
generation~\cite{li2022geometry}. Some studies~\cite{beetz2021generating, niederer2020creation, romero2021clinically, bonazzola2021image, danu2019deep} have adopted these approaches to generate virtual anatomy populations. For example, Beetz~\textit{et al.}~\cite{beetz2021generating} used a variational autoencoder (VAE)~\cite{kingma2013auto} to learn latent representations of cardiac biventricular anatomy represented as point clouds. They equipped the network with additional population-specific characteristics inputs to allow conditional synthesis of biventricular anatomies. Romero~\textit{et al.}~\cite{romero2021clinically} explored the efficiency of the generative adversarial network (GAN)~\cite{goodfellow2014generative}, trained on binary aorta masks. Danu~\textit{et al.}~\cite{danu2019deep} used a deep generative model to generate voxelised vessel surfaces and ensure compatibility between the unstructured representation (points/polygons) of vessel surfaces, and the structured domain required for the application of convolutional neural networks. Bonazzola~\textit{et al.}~\cite{bonazzola2021image} used a convolutional graph VAE to learn latent representations of image-derived 3D left ventricular meshes and used the learnt representations as surrogates for cardiac phenotypes in genome-wide association studies.

Although practical, all methods above were designed for single-part anatomies~\cite{romero2021clinically}, or require \textit{complete overlap} across all training samples in terms of the presence/absence of anatomical structures of interest in multipart shape assemblies~\cite{beetz2021generating}. Therefore, such techniques do not maximise the value of multiple disparate datasets with \textit{partial overlap} of anatomical structures, limiting the variability that can be synthesised in multipart shape assemblies (as the training population is limited to samples exhibiting \textit{complete overlap}). Recent advances in shape \textit{compositional} learning~\cite{luo2015learning, li2020learning, niemeyer2021giraffe, li2021editvae} within the computer vision domain look to address these limitations by leveraging complementary information available in disparate data sets to build rich generative models of multipart shape assemblies. For example, Luo~\textit{et al.}~\cite{luo2015learning} presented a compositional contour-based shape model incorporating multiple metrics to account for varying shape distortions or deformations. Li~\textit{et al.}~\cite{li2020learning} proposed to learn 3D shape synthesis through a two-stage framework including part generation and assembly, while GIRAFFE~\cite{niemeyer2021giraffe} modelled 3D scenes using compositional neural feature fields. 

\subsection{Contributions}
This study proposes a generative shape \textit{compositional} learning framework based on graph-convolutional neural networks to generate virtual cohorts of cardiovascular structures (represented as surface meshes/undirected graphs). We refer to the synthesised constructs as \textit{virtual chimaera} cohorts. The developed approach enables disparate data sets with partially overlapping anatomical structures to be used to learn a generative model of multipart shape assemblies. Although the proposed generative framework is demonstrated here in cardiovascular structures, it is generic by design and applicable to other multi-structure/organ ensembles. The key contributions of this study are as follows. 
\begin{enumerate}
	\item This is the first study to tackle the problem of combining data from different subjects with \textit{partially-overlapping} anatomical structures in a generative shape modelling framework to synthesise multipart shape assemblies representative of native anatomy. We refer to instances synthesised by the proposed approach as \textit{virtual chimaeras}. 
	\item Although generative shape compositional learning has previously been proposed within the computer vision domain, this is the first study to propose an approach for synthesising multipart assemblies of anatomical structures.
	\item Existing generative shape compositional learning approaches have used composition networks that predict rigid or affine transformations to compose the individual parts synthesised into multipart assemblies. We extend these approaches by including a nonrigid registration component in our composition network to reduce topological errors such as gaps or mesh intersections between individual parts (anatomical structures) of the shape assembly.
	\item We propose a novel self-supervised approach to train the shape composition network. The proposed approach uses the shared-boundary information between adjacent structures (i.e. nodes/vertices shared between surfaces of adjacent structures) to guide the training of the composition network. Existing approaches to learning generative shape composition have relied on \textit{strong supervision} for training the composition network, i.e., where (i) the ground truth spatial transformations required to compose multiple structures into a coherent multipart shape assembly spatially are known \textit{a priori}; or (ii) where the ground truth multipart shape assembly is available for each training sample. This precludes the use of partially overlapping data (the main motivation of this study) to train the shape composition network. The proposed self-supervised training scheme alleviates the need to access the spatial transformations of the ground truth or the complete multipart shape assembly \textit{a priori} and allows the composition network to be trained using partially overlapping data.
	
\end{enumerate}

Building generative models of multipart shape assemblies (such as the cardiovascular structures considered in this study, namely, four cardiac chambers and the root of the aortic vessel) is challenging due to the distinct variability in shape of each part, often the varying topology between the individual parts, and typically, the need for a suitable training set where all parts/structures of the shape assembly of interest are available for all samples (which precludes the combination of multiple datasets with partially overlapping anatomical structures). It is, therefore, desirable to formulate a generative framework that can effectively capture the variability in the shape of each part in an assembly, can accommodate varying topology across regions, and, finally, is not dependent on the availability of all aspects of the shape assembly, across all samples in the training set. To facilitate this, we explore two different generative approaches for learning part-aware latent representations of five cardiovascular structures (i.e. four cardiac chambers and the aortic root) and analyse them within the proposed compositional framework. These two approaches are called the \textit{independent generator} and the \textit{dependent generator}. The \textit{independent generator} comprises five independent graph-convolutional variational autoencoders (gcVAEs) corresponding to the five cardiovascular structures of interest. The \textit{dependent generator} is a novel graph-convolutional multichannel variational autoencoder (mcVAE). These are regarded minor contributions of the study in addition to the core contributions (1-4) listed above as - (i) convolutional mesh autoencoders (CoMA) \cite{ranjan2018generating} and gcVAE \cite{litany2018deformable} were previously proposed for different applications on 3D face reconstruction and deformable shape completion, respectively; and (ii) mcVAE was also previously proposed for learning shared latent representations of multimodal images \cite{antelmi2019sparse}. Our contributions in this regard are thus restricted to using these approaches within a novel shape compositional learning framework and formulating a graph-convolutional variant of the mcVAE. Finally, we also propose new metrics for evaluating the quality of synthesised virtual cardiac cohorts in terms of the overall variability in shape captured across the synthesised \textit{virtual chimaeras}, their anatomical plausibility, and their clinical relevance evaluated as ratios of standard cardiac volumetric indices, referred to as `clinical acceptance criteria'. 

\section{Methodology}

\begin{figure*}[!htbp]
	\centering
	\includegraphics[width=1\linewidth]{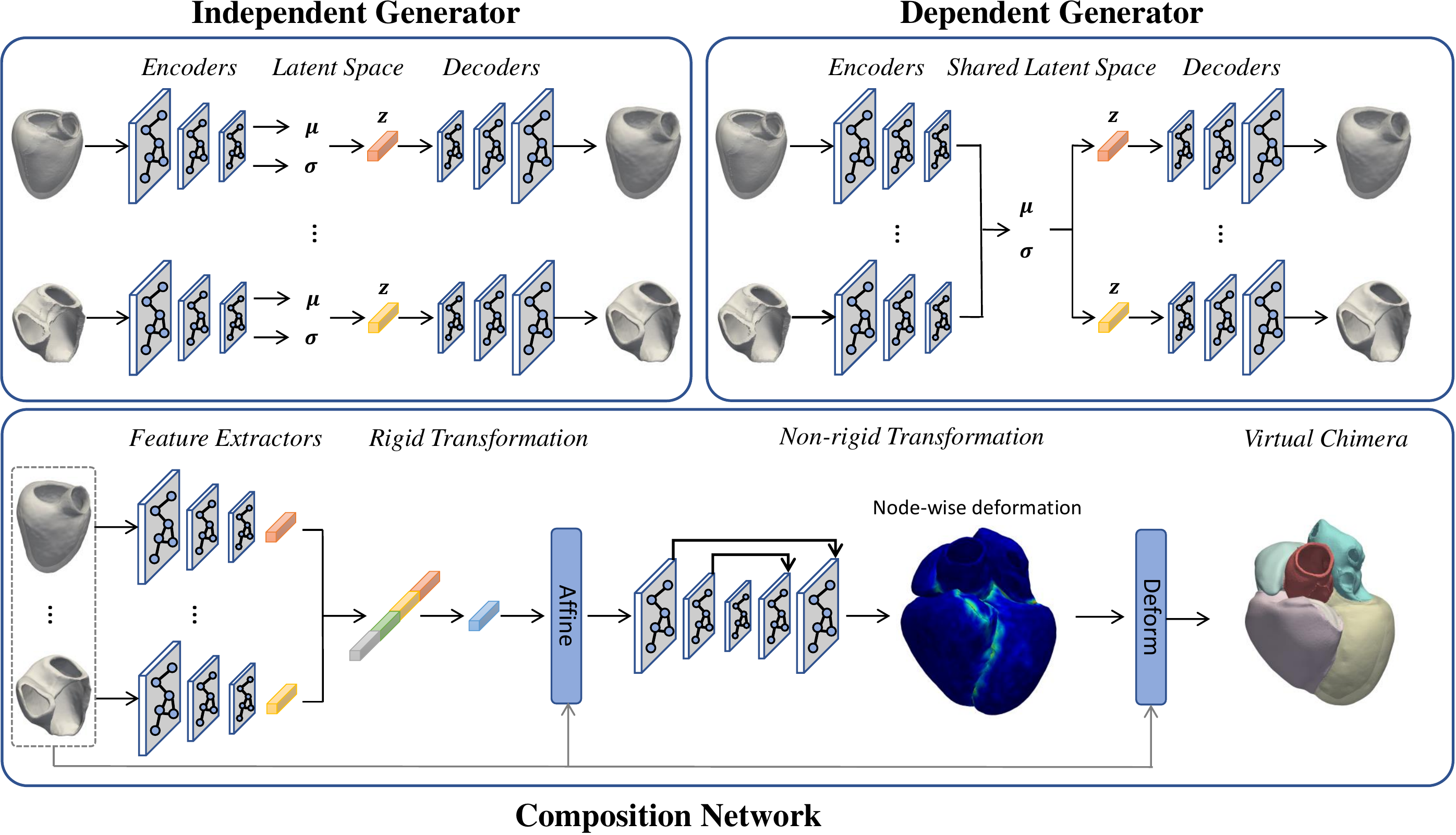}	
	\caption{Schematic illustration of our proposed shape compositional framework, which consists of a part-aware generator to learn the shape representations of each part and a composition network to perform the affine and nonrigid transformation to spatially compose the synthesised parts into a whole heart shape assembly.}
	\label{fig:framework}
\end{figure*}

The proposed deep compositional framework of shapes for synthesising virtual cardiac populations is described by the schematic shown in Fig.~\ref{fig:framework}. It comprises two modules: a \textit{part-aware generative model} and a \textit{composition network}. The part-aware generative model can be formulated in several ways, using traditional statistical/machine learning approaches (e.g. PCA, probabilistic PCA, etc.) or recent geometric deep learning approaches (e.g. gcVAEs). This study explores two generative schemes based on graph-convolutional neural networks: the \textit{independent generator} and the \textit{dependent generator}. The independent generator learns latent representations of each part individually, where each subgenerator is a gcVAE specific to its part. This provides flexibility in the overall generative framework since each gcVAE partwise is trained independently, allowing the use of \textit{partially overlapping patient data } to train each gcVAE. Additionally, as each gcVAE explicitly models the part-wise variability in shape observed across the training population(s), we argue that their combination within the proposed shape composition framework can synthesise virtual cardiac cohorts that capture a greater degree of variability in shape than afforded by using a single gcVAE (or similar model, e.g., PCA) to model all parts in the shape assembly jointly. 

In contrast, the dependent generator learns a shared latent representation across all parts of a shape assembly (that is, all four cardiac chambers and the aortic root in this study) using a graph-convolutional mcVAE network \cite{antelmi2019sparse}. The dependent generator models the joint likelihood of the observed data, i.e., of all parts in the shape assembly, as a product of the conditional likelihoods of each piece, conditioned on the other observed parts, the shared latent variables, and the model parameters. Estimating shared latent variables promotes anatomical plausibility in the shapes synthesised using the learnt latent representation, as the latter captures the covariation observed across multiple parts in the training population(s). As the dependent generator learns a shared latent representation of all elements in a shaping assembly, the model by design enables the conditional synthesis of one or multiple parts, given one or more other parts as inputs. Similarly, the dependent generator can also be trained using \textit{partially overlapping} patient data (i.e., when not all elements in a shape assembly of interest are available in all samples considered in the training population). We explore the independent and dependent generator within the proposed shape compositional framework to compare their relative merits in terms of the quality of the synthesised cardiac \textit{virtual chimaera} cohorts --- one is an approach designed to enhance the overall variability in the captured anatomical shape (as in the former); the other is designed to promote anatomical plausibility (as in the latter).

Cardiovascular structures synthesised using the part-aware generator (based on either the independent or dependent generator) are initially not spatially composed. Therefore, they do not represent complete, anatomically plausible hearts. A \textit{composition network} is required to facilitate this spatial composition. The composition network comprises a deep graph-convolutional neural network that first estimates 3D affine transformations to spatially organise each synthesised part/cardiovascular structure into a valid shape assembly. That is, the synthesised parts are transformed so that their relative positions and orientations to each other are consistent with native anatomy. Subsequently, the composition network refines the shape assembly by estimating a nonrigid transformation that locally deforms the nodes at interfaces between adjacent parts. This nonrigid transformation step is necessary as affine transformations alone could result in gaps and intersections between adjacent elements in a shape assembly, which is reduced through the former. The output of the composition network is a composed whole heart mesh/shape assembly, representing a \textit{ virtual heart chimera} instance (as illustrated in Fig.~\ref{fig:framework}).

\subsection{Part-Aware Generative Model}
As discussed, using the independent or dependent generator, the part-aware generative model enables the semantic part-wise synthesis of the cardiovascular structures of interest, namely the four cardiac chambers and the aortic root. We select graph convolution operations based on truncated Chebyshev polynomials~\cite{defferrard2016convolutional} and adopt mesh downsampling and upsampling operations as in CoMA~\cite{ranjan2018generating} to construct encoder-decoder networks for independent and dependent generators. As shown in Fig.~\ref{fig:framework}, each encoder-decoder network pair in the independent and dependent generator takes a triangular surface mesh of a part/cardiovascular structure as input and outputs the reconstructed surface mesh. Each mesh is represented by a list of 3D spatial coordinates of its vertices and an adjacency matrix defining vertex connectivity (i.e. edges of mesh triangles). Mesh downsampling and upsampling operations in the network help capture global and local shape contexts and are defined over a multi-resolution mesh hierarchy based on quadric edge collapses~\cite{garland1997surface}. The order is divided into five resolution levels. Features are learnt at each resolution level using a graph-convolution block with two constituent graph-convolution layers. In the encoder, the number of feature channels within each layer in each of the five blocks is 16, 32, 32, 64 and 64, respectively, with increasing network depth. Similarly, the same number of channels is used in reverse order for each block in the decoder.

\begin{figure}[!htbp]
	\centering
	\includegraphics[width=1\linewidth]{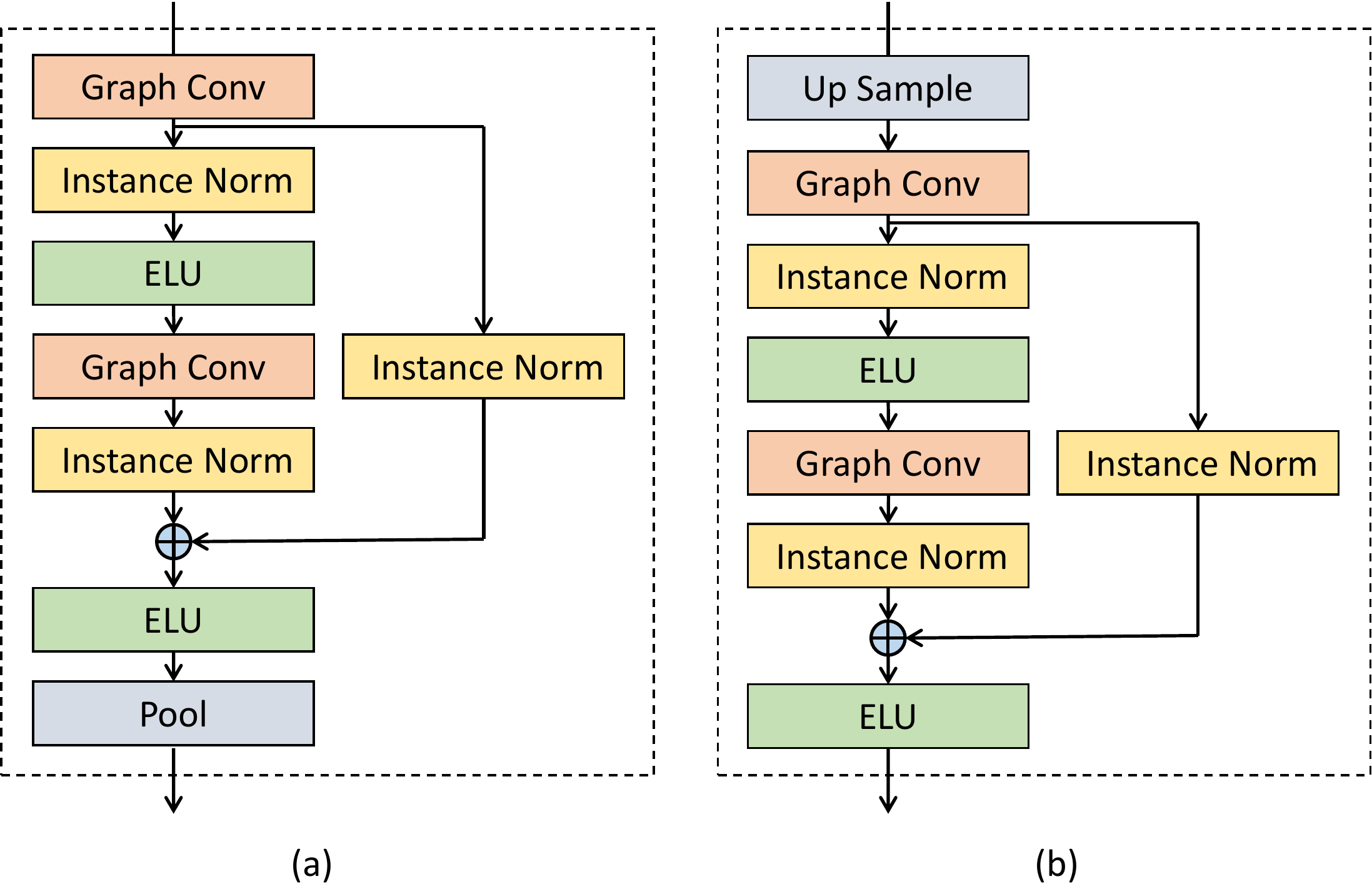}	
	\caption{Schematic illustration of the graph convolution blocks. (a) is the architecture of
		the residual graph convolution block in the encoder, and (b) is that in the decoder.}
	\label{fig:resGCN}
\end{figure}

To strengthen the efficiency of feature exchange between vertices~\cite{wang2018pixel2mesh} and mitigate the problem of vanishing gradients when training the network, we replace the basic graph convolution block used in~\cite{ranjan2018generating} with a residual block (shown in Fig.~\ref{fig:resGCN}). Each residual block comprises two graph convolution layers that extract features using the inputs provided to the union and a residual connection \cite{he2016deep} that combines the output of the first and second graph convolution layers through summation. Chebyshev graph convolution operations are used throughout this study due to their strictly localised filters, which enable learning of multiscale hierarchical patterns combined with mesh-pooling operations and low computational complexity. Each graph convolution layer is followed by an instance normalisation layer~\cite{ulyanov2016instance} and an exponential linear unit (ELU)~\cite{trottier2017parametric} for activation. Additionally, an instance normalisation layer is used in the residual branch to ensure similarity in feature statistics relative to the output of the second graph convolution layer in the block.

The part-aware generative model is trained in independent and dependent generators by optimising a composite loss function that combines three different losses. The independent and dependent generators are Bayesian models based on VAEs. The latent space from which the observed data $\mathbf{X}$ are generated can be discovered by approximating the posterior distribution of the latent variables using variational inference \cite{kingma2013auto}. This is achieved by optimising the evidence lower bound (ELBO) of the observed data concerning the networks' parameters, which can be formulated as a summation of the expected reconstruction error (i.e., expected negative log-likelihood of the data) and the Kullback-Leibler divergence of the approximate posterior distribution of the latent variables, from their assumed prior (as shown in \cite{kingma2013auto}). Based on this formulation of the ELBO, we construct our composite loss function to include a reconstruction loss $\mathcal{L}_{recon}$ based on the norm computed vertexwise $L_1$ between the predicted and original mesh vertices; the Kullback-Leibler divergence $\mathcal{L}_{KL}$ between the approximate posterior distribution of the latent variables $q(\mathbf{z} | \mathbf{X})$, and the prior distribution over the latent variables $p(\mathbf{z})$; and an additional regularisation loss $\mathcal{L}_{regular}$ that penalises outliers and encourages smooth surface reconstructions. A multivariate Gaussian prior $(p(\mathbf{z}) \sim \mathcal{N}(\mathbf{0},\mathbf{I}))$ with unit variance is used throughout this study for the independent and dependent generators. The key difference between the independent and dependent generator lies in the formulation of $\mathcal{L}_{recon}$, i.e. the expected negative log-likelihood of the observed data $\mathbf{X}$. The composite loss function used to train each part-aware gcVAE in the independent generator and the graph-convolutional mcVAE in the dependent generator is given by
\begin{equation}\label{equ-generation_loss}
	\mathcal{L} = \mathcal{L}_{recon} + \omega_0 \mathcal{L}_{KL} + \mathcal{L}_{regular}
\end{equation}
where, $\omega_0$ denotes the weight of the $\mathcal{L}_{KL}$ loss. The reconstruction loss $\mathcal{L}_{recon}$ for each gcVAE in the independent generator is computed as,
\begin{equation}
	\mathcal{L}_{recon} = \|\mathbf{X}_c - \mathbf{X}^{'}_c \|_1
\end{equation}
where, $X_c$ and $X^{'}_c$ denote the input and reconstructed mesh of part $c$. As the dependent generator is a graph-convolutional mcVAE that learns a shared latent representation across all $c = \lbrace 1...C \rbrace$ parts in a shape assembly, the joint data likelihood across all parts $\mathbf{X}_{c=1..C} \in \mathbf{X}$ can be expressed as $p(\mathbf{X}\mid\mathbf{z}) = \prod^{C}_{c=1} p(\mathbf{X}_c \mid \mathbf{z})$ by assuming that each part is conditionally independent of the others (conditioned on $\mathbf{z}$). Based on this assumption, the reconstruction loss $\mathcal{L}_{recon}$ is calculated as
\begin{equation}\label{equ-MCVAE}
	\mathcal{L}_{recon} = \sum^C_{c=1} \sum^C_{c'=1}\| \mathbf{X}_c - \mathbf{X}_{c|c'}^{'} \|_1
\end{equation}
where, $\mathbf{X}_{c|c'}^{'}$ is the reconstructed mesh of part $c$ conditioned on the shared latent space of all parts $c' = \lbrace 1...C \rbrace$. This formulation of the joint data likelihood based on mcVAE \cite{antelmi2019sparse} enables the reconstruction of a multipart shape assembly given a single part as input. Complete multipart shape assemblies can also be sampled from the shared latent space learnt.

Finally, $\mathcal{L}_{regular}$ is formulated identically for independent and dependent generators and contains two terms: a Laplacian smoothness loss and an edge length loss.  Laplacian loss encourages neighbouring vertices to move coherently, thus reducing self-intersections of the mesh and encouraging smoother surface reconstructions~\cite{nealen2006laplacian}. Given a vertex $\mathbf{v}$ and neighbouring vertices $\mathbf{v}_k$, where, $\mathbf{v}_k \in N(\mathbf{v})$, it is defined as:
\begin{equation}
	\begin{split}
		\label{equ-laplacian_loss}
		\mathcal{L}_{Laplacian} = \sum_{\mathbf{v}} \left (\mathbf{v} - \sum_{\mathbf{v}_k \in N(\mathbf{v})} \frac{1}{\| N(\mathbf{v}) \|} \mathbf{v}_k \right )
	\end{split}
\end{equation}

The edge length loss penalises spurious motion of vertices relative to $k$ neighbouring vertices and is defined as,
\begin{equation}
	\begin{split}
		\label{equ-edge_loss}
		\mathcal{L}_{edge} = \sum_{\mathbf{v}} \sum_{\mathbf{v}_k \in N(\mathbf{v})} \| \mathbf{v} - \mathbf{v}_k \|_{2}
	\end{split}
\end{equation}
Therefore, the overall regularisation loss ${L}_{regular}$ is defined as the weighted summation of both losses mentioned above and is given by,
\begin{equation}
	\begin{split}
		\label{equ-regularization}
		\mathcal{L}_{regular} = \omega_1 \mathcal{L}_{Laplacian} + \omega_2 \mathcal{L}_{edge}
	\end{split}
\end{equation}
The weights associated with the terms in the composite loss function $(\omega_0, \omega_1$ and $\omega_2)$ are hyperparameters tuned and set empirically.

\subsection{Spatial Composition Network}
Once the part-aware generative model is trained, using either the independent or the dependent generator, all parts in the shape assembly can be sampled from the model. Since each piece is generated separately, their spatial positions and orientations relative to one another are inconsistent with the native anatomy. Consequently, when initially assembled, synthesised parts may have significant intersections/overlaps with other features or large gaps with certain parts with which they are meant to share boundaries, as per native anatomy. Thus, learning the spatial relationships between elements in the assembly is imperative to estimate the spatial transformations necessary to compose the individual parts synthesised into a whole heart-shaped crowd consistent with native anatomy. Previous studies on learning shape composition in the computer vision domain~\cite{li2020learning,dubrovina2019composite} have relied on rigid registration for assembling synthesised individual parts using a dedicated spatial composition network and have predominantly applied such a generative framework to solid structures such as chairs, tables, planes, etc. However, due to the deformable nature of soft tissue, cardiovascular structures exhibit localised nonlinear variations in shape between patients.
Consequently, rigid/affine transformations alone are insufficient to accommodate such localised shape variations and to compose the individual parts synthesised into whole-heart shape assemblies consistent with native anatomy. Therefore, we propose a spatial composition network that estimates the affine and nonrigid transformations (see Fig.\ref{fig:framework}) and composes the cardiovascular structures synthesised by the part-aware generator into whole-heart shape assemblies. Furthermore, our approach differs distinctly from previous methods for learning shape composition~\cite{li2020learning,dubrovina2019composite} that rely on strong supervision to estimate the necessary spatial transformation. Instead, we propose a self-supervised learning scheme to calculate the desired affine and nonrigid transformations, driven by the shared vertices between adjacent structures in the shape assembly (known \textit{a priori}).

Figure~\ref{fig:framework} shows our composition network, which takes the generated shapes of each cardiovascular structure (that is, four cardiac chambers and the aortic root) as input and outputs the entire composed heart. Synthesised parts are first passed to the affine registration module of the composition network. This affine registration module comprises part-specific sub-networks utilising the same architecture as each encoder in the independent generator, which is trained simultaneously to extract features from the input shapes. These features are aggregated by concatenation and used to guide the estimation of the desired affine transformations (see Fig.\ref{fig:framework}), denoted $\mathcal{T}_{c=1...C}$, resulting in an initial coarsely composed whole-heart shape. The subscript indicates the $c$\textsuperscript{th} part in the shape assembly.

The estimated 3D affine transformations comprise eight parameters, including translation ($[T_x, T_y, T_z]$), scaling ($S$), and rotation ($[Q_1, Q_2, Q_3, Q_4]$ parameterised using quaternions). The loss function minimised to train the affine registration module of the composition network (in a self-supervised manner) and estimate the desired 3D affine transformations is given by
\begin{equation}
	\begin{split}
		\label{eq-rigid}
		\mathcal{L}_{affine} = \sum^{C}_{c=1} \|\mathbf{V}^{transf}_o - \mathcal{T}_c\mathbf{V}_c \|_1 
	\end{split}
\end{equation}
where $\mathcal{T}_c$ represents the affine transformation estimated and applied to part $c$ in the assembly; $\mathbf{V}_c$ denotes the vertices shared between part $c$ and all other $C-1$ parts in the assembly, where the latter (i.e. all other $C-1$ parts) are represented by subscript $o$; $\mathbf{V}^{transf}_o$ denotes the transformed vertices shared between all other $C-1$ parts in the assembly given the $c$\textsuperscript{th} part. Here, $\mathbf{V}^{transf}_o$ is constructed by concatenating shared vertices from all parts $C-1$ (given a part $c$ in the assembly), following an affine transformation. That is, $\mathbf{V}^{transf}_o = \lbrace \mathcal{T}_d \mathbf{V}_{d} \rbrace_{d=\lbrace 1...C-1\rbrace}$, where $\mathbf{V}_d$ denotes the vertices shared between part $d$ in the assembly and part $c$, for $d \neq c$.

Ensuring coherence at the boundaries between adjacent cardiovascular structures is essential to ensure anatomical plausibility in the synthesised \textit{virtual chimaera} cohorts and to prevent topological errors in whole heart shape assemblies. As rigid/affine registration alone is insufficient to prevent intersections and gaps between adjacent cardiovascular structures in the assembly, our composition network also comprises a nonrigid registration module, which deforms each part locally and refines the composition of the synthesised structures. The non-rigid registration module is a graph-convolutional neural network with a similar encoder-decoder architecture used for each gcVAE specific to each part in the independent generator, and additional skip connections~\cite{ronneberger2015u} aggregating the features in each encoder block with its corresponding decoder block at the same mesh resolution level. This module is trained as an autoencoder rather than a VAE. It inputs the coarsely aligned whole-heart shape of the preceding affine registration step. It estimates vertex-wise displacements to refine the spatial composition of the whole-heart shape assembly. The nonrigid registration module is also trained in a self-supervised manner (as with the affine registration module) by minimising a loss function given by
\begin{equation}
	\begin{split}
		\begin{aligned}
			\label{equ-assembling}
			\mathcal{L}_{non-rigid} = &\sum^C_{c=1} \|\mathbf{V}^{transf}_{o} - \mathcal{T}^{nr}_c\mathbf{V}_c \|_1 + \\ & \omega_3 \mathcal{L}_{Laplacian} + \omega_4 \| \mathcal{T}^{nr} \|_1
		\end{aligned}
	\end{split}
\end{equation}
where $\mathcal{T}^{nr}$ represents the nonrigid transformation, i.e. the vertex-wise displacements, estimated for all vertices of all parts in the coarsely aligned full heart shape output from the preceding affine composition step, and $\mathcal{T}^{nr}_c$ are the displacements estimated for the vertices shared between part $c$ (denoted $\mathbf{V}_c$) and all other $C-1$ parts in the assembly. The shared vertices of the latter (i.e., all other $C-1$ parts) are represented by subscript $o$ and $\mathbf{V}^{transf}_o$ denotes the transformed vertices shared between all other $C-1$ parts in the assembly and the $c$\textsuperscript{th} part. Here, $\mathbf{V}^{transf}_o$ is constructed by concatenating vertices shared between all $C-1$ parts and the $c$\textsuperscript{th} part in the assembly after deformation using estimated displacements in the vertices. That is, $\mathbf{V}^{transf}_o = \lbrace \mathcal{T}^{nr}_d \mathbf{V}_{d} \rbrace_{d=\lbrace 1...C-1\rbrace}$, where $\mathbf{V}_d$ denotes the vertices shared between part $d$ in the assembly and part $c$, for $d \neq c$. In equation~\ref{equ-assembling}, $\mathcal{L}_{Laplacian}$ represents the Laplacian loss (described in equation~\ref{equ-laplacian_loss}) used to regularise the estimated vertex-wise displacements, encouraging the latter to be similar for neighbouring vertices and resulting in smooth, localised deformations of each part in the shape assembly. The third term in the equation~\ref{equ-assembling} applies L1 normalisation to the vertex-wise displacements to encourage sparsity in the estimated vertex displacements and penalise the motion of vertices that are not in the vicinity of vertices shared between two adjacent regions. $\omega_3$ and $\omega_4$ are the empirically tuned regularisation parameters.

Following this two-step process, comprising an initial affine and a subsequent non-rigid registration step, the composition network is trained by optimising $\mathcal{L}_{affine}$ and $\mathcal{L}_{nonrigid}$, respectively, to spatially organise the cardiovascular structures synthesised by the part-aware generator into anatomically consistent \textit{ virtual heart chimeras}.

\section{Experiments Configuration}

\subsection{Datasets}
The generative shape compositional framework proposed in this study was trained and validated using a subset of cardiac cine-magnetic resonance (cine-MR) imaging data available from the UK BioBank (UKBB)~\cite{sudlow2015uk}. We created a cohort of 2,360 subject-specific meshes of the whole heart using the manual contours provided for the four cardiac chambers in a previous study \cite{petersen2017reference}. The subject-specific meshes of the entire heart were created by registering a high resolution heart atlas mesh of an earlier study \cite{rodero2021linking} with the manual contours. The cardiac atlas comprises the following cardiovascular structures: left/right ventricle (LV / RV), left/right atrium (LA / RA), the root of the aorta vessel, and four-valve planes (i.e. mitral, tricuspid, aortic and pulmonary valves). Additionally, vertices in the atlas mesh are shared between the following structures: LV-RV, LV-LA, LA-RA, RV-RA, LV-aorta, LA-aorta, and RA-aorta. Registration of the atlas mesh to each subject's manual contours results in a cohort of subject-specific meshes (i.e. undirected graphs) that share point correspondence, i.e. have the same number of vertices/nodes and share the same edge connectivity. This enables spectral convolutions in the graph-convolution layers used throughout the proposed generative-shape compositional framework (as fixed/common graph topology across all samples is a prerequisite for the former).
Additionally, the self-supervised learning scheme proposed in this study to train the composition network is driven by the shared vertex information available for adjacent parts/structures in the cardiac atlas mesh (which is automatically propagated to all subject-specific meshes via registration). Atlas-to-contour registration was achieved using a combination of the point set registration algorithm (viz. Generalised coherent point drift) developed previously by our group \cite{ravikumar2017generalised} to establish soft correspondences between atlas and subject-specific contours and subsequent thine-plate-spline-based warping of the atlas mesh to each subject's outlines. Further details on the registration process used to create the cohort of subject-specific meshes used in this study are available in \cite{xia2022}. We randomly split the general cohort of whole heart meshes from 2360 subjects into 422/59/1879 for training, validation, and testing, respectively. We explore two distinct scenarios when training the proposed generative shape compositional framework. These include (a) \textit{complete overlap} in the data across all training samples, that is, where the data of all subjects included in the training set contain all cardiovascular structures in the whole heart shape assembly; and (b) \textit{partial overlap} in the data across training samples, where 300 subjects have only one cardiovascular structure, resulting in 60 samples for each of the five structures of interest (ie LV, RV, LA, RA and aorta), while the remaining 122 subjects in the training set are considered to have complete assembly of all heart shape. Throughout this study, for each subject, we only use meshes representative of one-time point in the cardiac cycle, namely, at end-diastole. The samples in the resulting training, validation, and test sets are used to train and evaluate the proposed approach throughout the study.

In this study, we evaluated our study based on partially overlapping scenarios simulated in the UKBB datasets. However, such scenarios can also be created based on multiple public and private datasets by registering the template mesh with patient-specific segmentations or meshes from the latter, as explained above. Examples of such data sets can be found in MedShapeNet~\cite{li2023medshapenet}, MM-WHS~\cite{zhuang2019evaluation}, Sunny Brook Cardiac Data~\cite{radau2009evaluation}, M$\&$Ms challenge~\cite{campello2021multi}, and Reprise III clinical trial data (Private)~\cite{feldman2018effect}, among others.

\subsection{Implementation Details}
The proposed approach was implemented using PyTorch~\cite{paszke2019pytorch} on a PC with an NVIDIA RTX 2080Ti GPU. We trained part-aware generative models (i.e., both independent and dependent generator) using the Adam optimiser~\cite{kingma2014adam} with an initial learning rate of $1e^{-4}$ and batch size of 16 for 600 epochs. The order of the Chebyshev convolution polynomial was set to 6. The latent dimensions of each part-specific subnetwork in the independent generator were set to 16, 12, 16, 12, and 8 for the LV, RV, LA, RA, and aorta, respectively. However, the latent dimension of the dependent generator was set to 60, with 12 components of the latent vector dedicated to modelling the shape variability observed in each of the five cardiovascular structures of interest, namely, LV, RV, LA, RA, and aorta. We set the resampling factors for learning shape features across a multi-resolution mesh hierarchy to: $[4, 4, 4, 6, 6]$ for the LV and RV, $[4, 4, 4, 4, 5]$ for the LA, $[4, 4, 4, 4, 6]$ for the RA, and $[4, 4, 4, 4, 4]$ for the aorta. These down- and up-sampling factors were chosen empirically for the five graph-convolution blocks in the encoder and decoder networks, respectively, in independent and dependent generators. During the training of the independent and dependent generator, a warm-up strategy was adopted to improve stability and prevent mode collapse in the learnt posterior distribution over the latent variables. This is achieved by initially training the partwise gcVAEs in the independent generator and the mcVAE in the dependent generator, with the weight of KL loss term (i.e. $\omega_0$ in equation~\ref{equ-generation_loss}) set to zero for $300$ epochs (i.e., they are trained as plain autoencoders). The learnt weights initialise the subsequent training step for the independent and dependent generator, where the importance of the KL loss term is initially set to a small value, ie $1e^{-6}$, and, subsequently, increased each epoch by multiplying by a factor of $1.25$, up to a maximum value of $1e^{-4}$. The weight of the other regularisation terms in the composite loss function (see Equation~\ref{equ-regularization}), namely $\omega_1$ and $\omega_2$, was set to $8$ and $1$. The weights for the regularisation terms in the self-supervised registration loss function (refer to equation~\ref{equ-assembling}) used to train the composition network, namely, $\omega_3$ and $\omega_4$, were set to $8$ and $5e^{-7}$, respectively. 

The hyperparameters of the network architecture and the optimiser used for the part-aware generator were retained for the composition network, except for the initial learning rate and batch size, which were set to $5e^{-4}$ and $4$, respectively. We randomly sampled individual part shapes from different patients to train the composition network. This helped to demonstrate that the proposed composition network can be trained with \textit{partially overlapping} data. 

Specifically, the composition network was trained in two stages. We first trained the model with only $\mathcal{L}_{affine}$ to ensure that the model learnt the relative pose of the parts. Subsequently, we fixed the parameters of the affine part in the network and trained the remaining parameters with $\mathcal{L}_{nonrigid}$. Additionally, we adopted a curriculum learning strategy that feeds data to the model in an easy-to-hard manner (real-to-synthetic), stabilising the optimisation. The composition network was first pre-trained using parts sampled from the original patient-specific cardiac meshes. The data splits used to train and evaluate part-aware generative models were kept the same for the composition network to ensure a fair evaluation. Subsequently, we fine-tuned the composition network for 200 epochs using parts synthesised by the IG-CO part-aware generative model. The data set of the synthesised parts comprised 2000 samples, randomly split into 1600/200/200 for training, validation, and testing, respectively.

All hyperparameters associated with the part-aware generative models and the composition network were tuned based on the validation set. The performance of the proposed generative shape composition framework was evaluated in the holdout test set, using several metrics designed to assess generalisability, specificity, and plausibility in terms of critical clinical cardiac indices, evaluated across synthesised \textit{virtual chimaera} cohorts (relative to the actual population of the UK Biobank). 

\subsection{Generating Virtual Chimaera Cohorts}
When developing/choosing generative shape models for any given application and sampling strategies used to synthesise virtual cohorts of anatomy from the former, it is typical to make design choices that achieve a balance/trade-off between the variability in shape captured (relative to target/natural patient populations), and the anatomical plausibility of the instances synthesised in the virtual cohorts. Balance of shape variability and anatomical likelihood in synthesised virtual cohorts is essential for \textit{in-silico} trials as cohorts with high variability may contain unrealistic shapes, unrepresentative of native anatomy/variations occurring naturally. On the contrary, cohorts synthesised with strict plausibility restrictions may not be expressive in the range of shape variations they represent, limiting the anatomical envelope that can be explored in subsequent \textit{ in vitro} trials. In \cite{romero2021clinically}, a detailed quantitative evaluation of the benefits and drawbacks of applying different sampling strategies to PCA-based shape models was explored by synthesising virtual cohorts of the aortic root and vessel. Romero et al. concluded that uniform sampling in the learnt latent space/principal subspace of a shape model yields cohorts with more significant shape variability, while sampling from unit Gaussian distributions to generate latent vectors representative of virtual shape instances (for PCA-based shape models) ensures greater plausibility in the synthesised virtual cohorts. Most existing studies generate new samples by sampling from the unit Gaussian distribution~\cite{ranjan2018generating, zhou2020fully}. In this study, we favour the synthesis of cohorts that maximises the variability in shape captured for each cardiac structure of interest and, therefore, opt for a uniform sampling strategy to synthesise the \textit{virtual chimaeras}. The rationale was to assess whether virtual cohorts synthesised using the different generative shape models investigated in this study could capture the variability in cardiac shape and associated clinically relevant cardiac indices observed in a natural (and unseen) population. Specifically, using the generative shape models proposed in this study (i.e., the independent and dependent generator), we synthesise cardiac \textit{ virtual chimaera} cohorts by uniformly sampling each latent variable within the mean $\pm 2$ standard deviation interval, where the means and standard deviations for each latent variable are learnt from the training data.
Similarly, for fair comparison, virtual cohorts are synthesised using the PCA-based shape model by sampling coordinates in the low-dimensional principal subspace uniformly in the interval defined $[-2\sqrt{\lambda}_i, +2\sqrt{\lambda}_i]$, where, $\lambda_i$ denotes the eigenvalue of the $i$\textsuperscript{th} principal component (i.e. $i$\textsuperscript{th} eigenvector that spans the principal subspace). The PCA-based shape model used throughout this study retained 24 principal components, explaining $95\%$ the variation in cardiac shapes observed in the training set used for all models. Additionally, when sampling latent vectors/coordinates in principal subspace, two standard deviations about the mean were used throughout, as we experimentally verified that sampling with three standard deviations or more resulted in a large number of irregular/implausible shapes in the synthesised virtual cohorts. 

\subsection{Evaluation Criteria}
We evaluated the proposed framework for the composition of synthesised shapes in terms of the following criteria: generalisability, which measures the ability of trained shape models to reconstruct unseen cardiac conditions and thus assesses the variability of shape captured by the learnt latent representations; specificity, which sets anatomical plausibility in the cardiac virtual cohorts synthesised; and clinical relevance. Generalisation errors were evaluated using the Euclidean distance (ED) and F1-score~\cite{knapitsch2017tanks, fan2017point}. Precisely, unseen test shapes were reconstructed using trained shape models investigated in this study, i.e. PCA, IG-CO, IG-PO, DG-CO, and DG-PO. Subsequently, the ED and F1 scores were evaluated between the original and reconstructed test shapes. Similarly, specificity errors were quantified as the ED between each sample in the synthesised virtual populations and its nearest neighbour (in terms of ED) in the actual unseen population of cardiac shapes. ED was evaluated as the vertex-to-vertex Euclidean distance between two cardiac bodies. 
The F1 score computes the harmonic mean of the precision and recall of the reconstruction. The accuracy is the fraction of predicted points in the reconstruction with the nearest neighbour in the ground truth within a certain distance. At the same time, recall is the fraction of correct points in the ground truth with the nearest neighbour in the reconstruction within the same threshold.

Furthermore, the registration errors incurred by the composition network were quantified using the Hausdorff distance (HD). Specifically, HD measured the distance between the vertices shared across all pairs of adjacent cardiac structures at their shared boundaries, following affine and non-rigid composition.

\section{Results}

We conducted several experiments to evaluate the performance of the generative shape models proposed in this study and compared them with each other and PCA. This section summarises the results obtained regarding generalisation and specificity errors for the methods investigated and the population-level quality of their synthesised virtual heart populations, assessed in terms of critical clinical cardiac indices. 

\subsection{Generalisability}
\begin{figure*}[!htbp]
	\centering
	\includegraphics[width=1\linewidth]{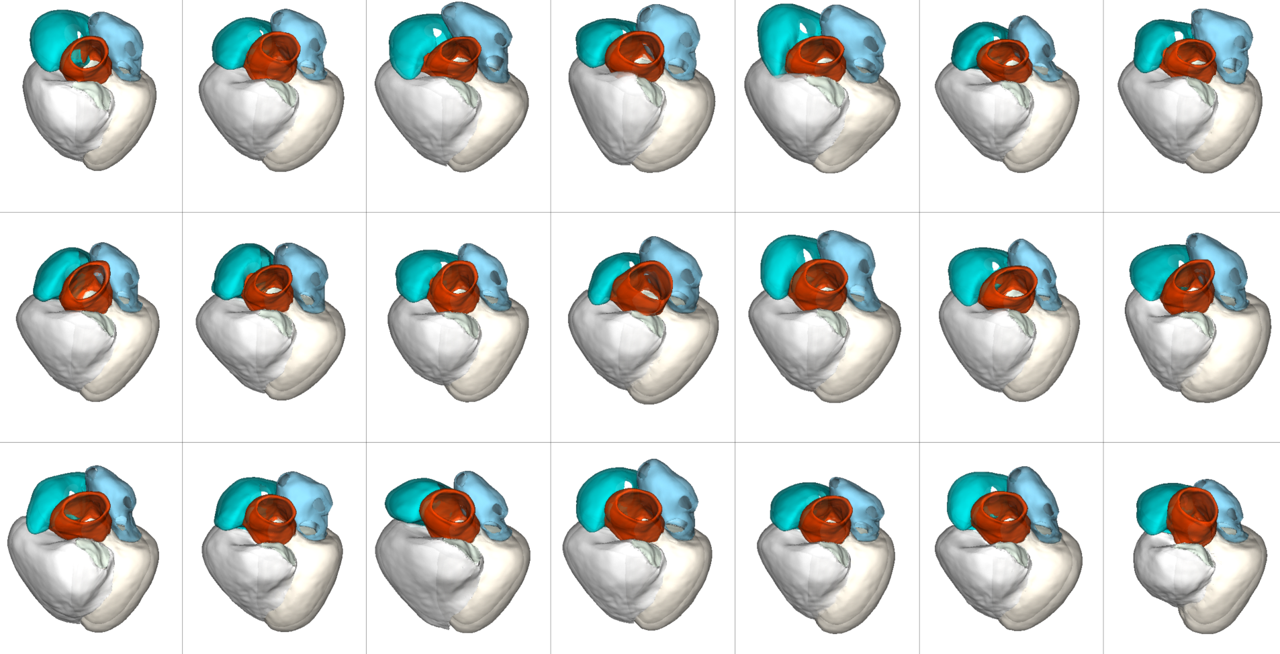}	
	\caption{Examples of \textit{virtual chimaeras} generated by the independent generator trained with \textit{complete data} (IG-CO). Each component of the latent vector was randomly sampled from the posterior Gaussian distribution learnt from the training dataset.}
	\label{fig:virtual_chimaera}
\end{figure*}

A model with good generalisability can capture the variability of the seen (training) data and generalise to or explain the unseen (testing) data. The generalisability of a generative/statistical shape model can be assessed in terms of the error incurred when reconstructing unseen test data, thus evaluating its ability to explain unseen shapes and providing information on the overall variability in shape captured by the model\cite{davies2009building}. Table~\ref{tab:generalization_error} summarises the generalisation errors of all methods investigated in this study for the five cardiovascular structures of interest. We observe that the IG models significantly outperform PCA for three of the five structures of interest (i.e. LV, LA, and RA). Of particular importance to note is that IG-PO, which was trained with missing/partially overlapping data, also outperformed PCA, which, conversely, was introduced with complete data.
Additionally, both IG models consistently outperformed the DG models in generalisation errors across all five structures. This indicates that by learning an independent part-specific latent space, the IG models can capture a greater degree of shape variability for each structure of interest than afforded by PCA and correspondingly, can synthesise more diverse virtual heart populations (in terms of shape) than the latter. Examples of virtual heart chimaeras generated by spatially composing (using the composition network) individual parts/structures sampled from the trained IG-CO model are shown in Figure~\ref{fig:virtual_chimaera} (examples of virtual hearts synthesised by all investigated methods are included in the supplementary material). 

\begin{table*}[!htbp]
	%\scriptsize
	\centering
	\caption{Generalisation errors of the proposed independent/dependent generator trained on complete/partially overlapping dataset, and the PCA model, on the holdout test dataset (mean$\pm$std). The \textbf{bold} results indicate statistically significant improvements over the PCA model.}
	\label{tab:generalization_error}
	\setlength{\tabcolsep}{0.75mm}{
		\begin{tabular}{ccccccccccc}
			\hline
			\hline
			\multicolumn{1}{c}{\multirow{3}{*}{Part}} &\multicolumn{2}{c}{\multirow{2}{*}{PCA~\cite{frangi2002automatic}}} &\multicolumn{4}{c}{Dependent Generator} &\multicolumn{4}{c}{Independent Generator} \\ \cline{4-11} 
			
			\multicolumn{1}{l}{} &\multicolumn{2}{c}{} &\multicolumn{2}{c}{Complete~\cite{antelmi2019sparse}} &\multicolumn{2}{c}{Partial} &\multicolumn{2}{c}{Complete~\cite{ranjan2018generating}}&\multicolumn{2}{c}{Partial} \\ \cline{2-11}
			
			\multicolumn{1}{l}{} &ED &F1 &ED &F1 &ED &F1 &ED &F1 &ED &F1 \\ \hline
			LV &2.37$\pm$0.71 &0.71$\pm$0.11 &\textbf{1.78$\pm$0.38} &\textbf{0.72$\pm$0.66} &\textbf{1.92$\pm$0.39} &0.70$\pm$0.11 &\textbf{0.88$\pm$0.17} &\textbf{0.95$\pm$0.05} &\textbf{1.43$\pm$0.25} &\textbf{0.82$\pm$0.08} \\
			
			RV &1.74$\pm$0.39 &0.82$\pm$0.07 &2.68$\pm$0.55 &0.66$\pm$0.08 &2.76$\pm$0.61 &0.65$\pm$0.09 &1.84$\pm$0.62 &0.75$\pm$0.15 &2.22$\pm$0.55 &0.71$\pm$0.11 \\
			
			LA &2.10$\pm$0.51 &0.82$\pm$0.07 &2.70$\pm$0.62 &0.71$\pm$0.09 &2.51$\pm$0.57 &0.73$\pm$0.09 &\textbf{1.47$\pm$0.41} &\textbf{0.89$\pm$0.08} &\textbf{1.84$\pm$0.50} &\textbf{0.83$\pm$0.09} \\
			
			RA &1.43$\pm$0.37 &0.92$\pm$0.05 &1.99$\pm$0.61 &0.82$\pm$0.10 &2.15$\pm$0.62 &0.80$\pm$0.10 &\textbf{1.21$\pm$0.40} &\textbf{0.93$\pm$0.09} &\textbf{1.43$\pm$0.49} &0.90$\pm$0.09 \\
			
			Aorta &1.40$\pm$0.40 &0.92$\pm$0.06 &1.98$\pm$0.55 &0.80$\pm$0.10 &2.18$\pm$0.67 &0.77$\pm$0.12 &1.51$\pm$0.48 &0.87$\pm$0.11 &1.71$\pm$0.45 &0.83$\pm$0.10  \\
			
			Full Heart &2.30$\pm$0.74&0.77$\pm$0.08  &3.53$\pm$0.76 &0.66$\pm$0.07 &3.48$\pm$0.71 &0.66$\pm$0.06 &2.97$\pm$0.73 &0.71$\pm$0.08 &3.18$\pm$0.75 &0.69$\pm$0.08 \\
			
			\hline \hline
	\end{tabular}}
\end{table*}

\subsection{Specificity}
Specificity errors are used to assess the anatomical plausibility of quantitatively synthesised virtual heart populations. This is done by evaluating the distance of each generated sample in the virtual population to the closest/most similar shape in the real population ~\cite{davies2009building}. Table~\ref{tab:specificity} summarises the specificity errors obtained for the virtual heart populations synthesised using each investigated model. Errors are quantified for each of the five structures of interest individually and for the whole heart shape assemblies. We observe that DG-CO achieves the highest specificity (i.e., lower specificity errors indicate greater plausibility) of all the methods investigated across all structures except the aorta. While DG-PO, which was trained with \textit{missing parts/partially overlapping} data, is comparable to PCA in all structures. By learning a joint latent representation across all parts in the whole heart shape assemblies, the dependent generator models constrain the posterior distribution of the latent variables to explain shared variation across multiple parts. Intuitively, this enforces greater plausibility in the whole heart shape instances generated from the trained model.

In contrast, the independent generator model IG-CO, trained using complete data, has lower specificity (i.e., higher specificity errors) than PCA and the dependent generator models for the ventricles, but is comparable to the latter for the atria. As the independent generator models learn distinct latent representations for each part/structure, covariation in shape across multiple parts is not captured. Consequently, the known partwise latent spaces focus on maximally capturing the variability in shape of each part, at the cost of plausibility in the instances generated.  

\begin{table*}[!htbp]
	\centering
	\caption{specificity errors of the virtual populations from different models (mean $\pm$ std). The \textbf{bold} results indicate that specificity is significantly better than or has no significant difference to the PCA model.}
	\label{tab:specificity}
	\begin{tabular}{ccccccc}
		\hline
		\hline
		Methods & Full Heart  & LV & RV & LA & RA & aorta\\ \hline
		PCA~\cite{frangi2002automatic}   &3.262$\pm$0.007  &2.211$\pm$0.006 &2.687$\pm$0.009 &2.688$\pm$0.010 &3.054$\pm$0.011 &2.217$\pm$0.008    \\
		DG-CO\cite{antelmi2019sparse}&\textbf{3.160$\pm$0.007} &\textbf{1.974$\pm$0.006} &\textbf{2.647$\pm$0.006} &\textbf{2.524$\pm$0.008} &\textbf{2.958$\pm$0.013} &2.399$\pm$0.013   \\
		DG-PO &\textbf{3.325$\pm$0.017} &\textbf{2.219$\pm$0.025} &\textbf{2.640$\pm$0.020} &\textbf{2.598$\pm$0.009} &\textbf{3.058$\pm$0.020} &2.254$\pm$0.015     \\
		IG-CO~\cite{ranjan2018generating} &3.780$\pm$0.073 &2.558$\pm$0.064 &3.668$\pm$0.122 &\textbf{2.731$\pm$0.044} &\textbf{3.094$\pm$0.048} &2.555$\pm$0.053     \\
		IG-PO &3.353$\pm$0.006 &\textbf{2.181$\pm$0.005} &\textbf{2.653$\pm$0.008} &\textbf{2.712$\pm$0.007} &3.150$\pm$0.011 &2.312$\pm$0.008     \\ \hline \hline
	\end{tabular}
\end{table*}

\subsection{Evaluation of clinical relevance}
\label{clinrev}
Based on the evaluation criteria proposed in~\cite{romero2021clinically} to assess virtual aorta cohorts synthesised by generative shape models, we evaluate the clinical relevance of virtual heart cohorts synthesised using the models investigated in this study. Specifically, we define the clinical acceptance rate $(\mathcal{A})$ as the percentage of synthesised samples in the virtual cohorts whose cardiac indices, namely, LV volume, RV volume, LA volume, and RA volume, are within a $90\%$ confidence interval of the distribution of these indices, observed in the UK Biobank population used in this study. As the distributions of these indices were verified to be non-Gaussian using the Kolmogorov-Smirnov test, we rely on Chebyshev's inequality to define the $95\%$ confidence interval based on the corresponding mean $(\mu)$, variance $(\sigma^2)$ and the mode $(M)$ observed in a real-world population. According to Chebyshev's inequality, intervals defined by $M \pm 2B$, where $B = \sqrt{\sigma^2 + (M-\mu)^2}$, contain at least $90\%$ of the area under the corresponding probability density functions \cite{amidan2005data}. 

\begin{table*}[!htbp]
	\centering
	\caption{Clinical acceptance rate $\mathcal{A} (\%)$ for cardiac virtual cohorts synthesised using all methods investigated.}
	\label{tab:clinical-accept}
	\begin{tabular}{c c c c c c c c c c}
		\hline
		\hline
		\multirow{2}{*}{Cardiac Indices} &\multirow{2}{*}{$\mu$} &\multirow{2}{*}{$\sigma$} & \multirow{2}{*}{$M$} & \multirow{2}{*}{$M \pm 2B$} &\multirow{2}{*}{PCA~\cite{frangi2002automatic}} & \multicolumn{2}{c}{Dependent Generator} & \multicolumn{2}{c}{Independent Generator}\\
		\cline{7-10}
		\multicolumn{6}{l}{}
		&Complete~\cite{antelmi2019sparse} &Partial &Complete~\cite{ranjan2018generating} &Partial \\
		\hline
		LV Volume
		&131.94 &20.39 &127.06 &[85.12, 168.99] &98.05 &100 &96.95 &98.15 &99.90 \\
		RV Volume
		&116.75 &20.66 &118.19 &[76.77, 159.61] &93.50 &99.50 &98.30 &90.50 &97.30\\
		LA Volume
		&34.98 &9.78 &30.69 &[9.33, 52.04] &84.60 &98.90 &98.50 &94.85 &99.65\\
		RA Volume
		&52.39 &10.87 &47.93 &[24.43, 71.43] &87 &95.40 &97.25 &96.65 &99.10\\
		\hline
		\hline
	\end{tabular}
\end{table*}

\begin{figure*}[!htbp]
	\centering
	\includegraphics[width=1\linewidth]{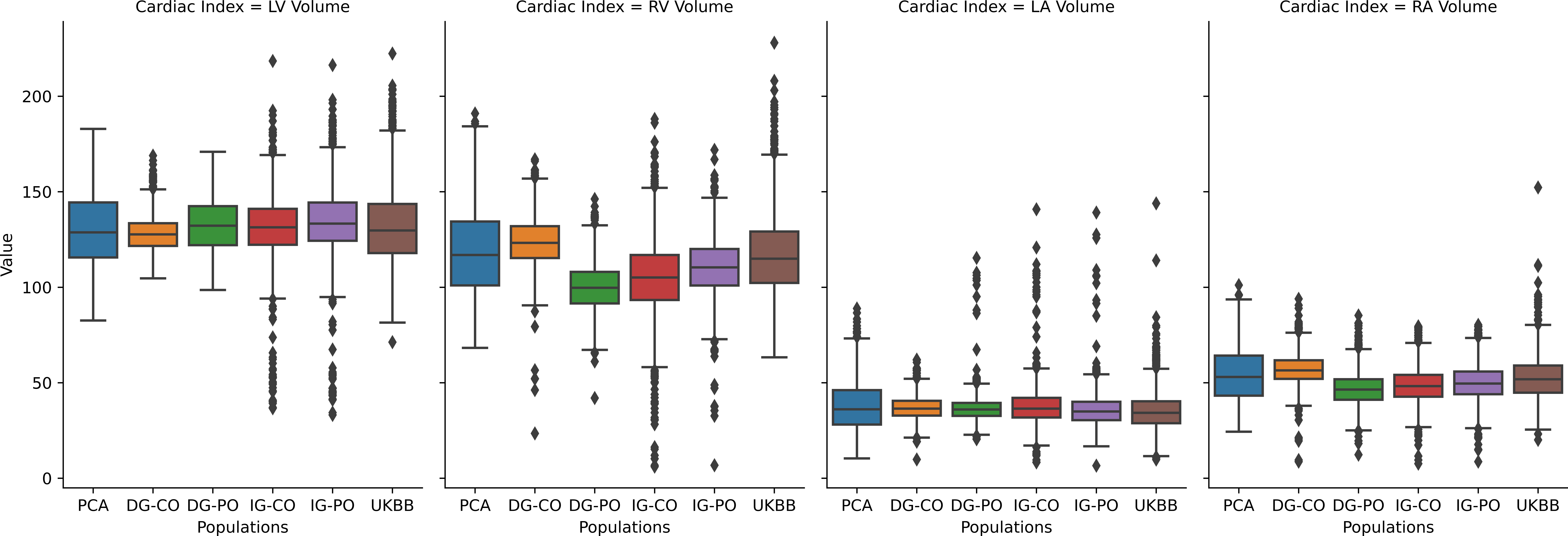}	
	\caption{Box plots compare the median and interquartile range for all four cardiac indices of interest across the synthesised virtual cohorts and the UK Biobank population. Boxplots for each synthesised virtual cohort are labelled according to the corresponding method used: PCA, IG-CO, IG-PO, DG-CO and DG-PO.}
	\label{fig:cardiac-indexes}
\end{figure*}

The clinical acceptance rate $\mathcal{A}$ is used as an additional metric to assess the anatomical plausibility of the synthesised virtual cohorts. It is motivated by the need to preserve clinically relevant cardiac volumetric indices in the synthesised cohorts (relative to the UK Biobank population reference). Table~\ref{tab:clinical-accept} summarises the statistical description of each cardiac index in the entire population and $\mathcal{A}$ calculated for all four cardiac chambers in the virtual cohorts synthesised. The dependent generator models (DG-CO and DG-PO) consistently obtain higher acceptance rates across all cardiac indices than the PCA model, consistent with the specificity errors summarised in Table~\ref{tab:specificity}. Although specificity errors indicate that independent generator models synthesise cardiac chamber shapes that are less plausible than PCA in some cases, the clinical acceptance rates estimated for IG-CO and IG-PO are consistently higher than PCA in all cardiac indices (refer to Table~\ref{tab:clinical-accept}). This indicates that the independent and dependent generator models proposed in this study, when trained with \textit{complete} and \textit{partially overlapping} data, provide better fidelity in preserving the distributions of clinically relevant cardiac indices in the virtual cohorts synthesised relative to the UK Biobank population than PCA. The range of values estimated for all four cardiac indexes of interest, namely, LV-volume, RV-volume, LA-volume and RA-volume, across all synthesised virtual cohorts and the actual UK Biobank population considered in this study, are summarised as box plots in Figure~\ref{fig:cardiac-indexes}.

\subsection{Evaluation of nonrigid spatial composition}
An essential contribution to this study is formulating a nonrigid self-supervised composition network to spatially compose the synthesised cardiovascular structures into coherent whole-heart shape assemblies (that is, \textit{ virtual heart chimaeras}). While previous approaches to generative shape composition learning have proposed rigid/affine composition networks, this is the first study to adopt nonrigid registration for the same (in addition to estimating an initial affine transformation). To demonstrate the benefits of non-rigid registration of the composition network, we compared the quality of the composed \textit{virtual heart chimaeras} obtained using only affine registration with those obtained by affine and non-rigid registration. Data from the holdout test set and samples from the virtual cohort synthesised using IG-CO were used in this experiment. Table~\ref{tab:share_vertex} summarises the Hausdorff distance (HD) between the shared vertices of all seven pairs of adjacent structures in the whole heart shape assemblies with shared boundaries, following composition with (w/) and without (w/o) nonrigid registration. These results indicate that the nonrigid spatial composition network consistently outperforms its affine counterpart across all adjacent structures, achieving significantly lower HD errors (reported as mean$\pm$std calculated across all samples $N$ used in the experiment in Table~\ref{tab:share_vertex}). Thus, non-rigid registration is a crucial component for synthesising \textit{virtual heart chimaeras} using the proposed framework, as it improves coherence at boundaries between adjacent structures and reduces topological defects compared to affine registration alone.

\begin{table}[!htbp]
	\centering
	\caption{Hausdorff distances (mean$\pm$std) estimated between shared vertices of all pairs of adjacent structures with shared boundaries in the whole heart shape assemblies. Whole heart shape assemblies were composed using the spatial composition network with (w/) and without (w/o) nonrigid registration, based on data from (a) the holdout test dataset, which was reconstructed using IG-CO; and (b) parts/structures synthesised by the trained IG-CO model. }
	\label{tab:share_vertex}
	\setlength{\tabcolsep}{0.75mm}{
		\begin{tabular}{c c c c c}
			\hline
			\hline
			\multirow{2}{*}{Adjacent Structures}   & \multicolumn{2}{c}{Reconstructed (N=1879)}  & \multicolumn{2}{c}{Synthesised (N=200)}  \\
			\cline{2-5}
			&w/o (mm) &w/ (mm)
			&w/o (mm) &w/ (mm)\\
			\hline
			LV-RV
			&6.30$\pm$1.61 &1.69$\pm$0.42 
			&7.90$\pm$5.87 &2.53$\pm$3.44 \\
			LV-LA
			&5.09$\pm$1.35 &1.51$\pm$0.32 
			&6.17$\pm$6.06 &1.94$\pm$2.27 \\
			LA-RA
			&4.26$\pm$1.22 &1.26$\pm$0.28 
			&4.78$\pm$1.37 &1.40$\pm$0.54 \\
			RV-RA
			&5.74$\pm$1.67 &1.39$\pm$0.29 
			&6.55$\pm$2.99 &1.67$\pm$1.24 \\
			LV-aorta
			&4.49$\pm$1.24 &1.40$\pm$0.28 
			&5.54$\pm$6.24 &2.15$\pm$3.92 \\
			LA-aorta
			&2.97$\pm$1.04 &0.88$\pm$0.20 
			&3.20$\pm$1.22 &1.08$\pm$0.94 \\
			RA-aorta
			&5.09$\pm$1.43 &1.66$\pm$0.43 
			&6.25$\pm$2.11 &2.23$\pm$0.94 \\
			
			\hline
			\hline
	\end{tabular}}
\end{table}

\section{Discussion}
This study presents a new shape compositional framework to synthesise virtual heart populations. While several previous studies have proposed generative shape models for synthesising virtual populations of multipart anatomical shapes, they have all relied on the availability of training data with \textit{complete overlap} across all samples, i.e. all structures/parts of interest being available across all samples in the training set. Consequently, existing approaches are not designed to combine the shape information available in \textit{partially overlapping} data, i.e., individual structures/parts from different samples/patients, and synthesise complete multipart shape assemblies. Leveraging \textit{partially overlapping} data is especially useful in scenarios where combining information from disparate multimodal data sets is necessary to enrich the training set and capture the desired variability in the shape of multipart shape assemblies. The two generative models introduced within the framework for shape composition proposed in this study, namely, the independent and dependent generators, were shown to synthesise plausible whole-heart shape assemblies using \textit{partially overlapping} training data. The specificity of the IG-PO and DG-PO models was comparable to or better than that of PCA, where the latter was trained with \textit{complete} data. In addition to synthesising plausible shape instances, generative shape models should generalise well to unseen shape instances, indicating the variability in shape (observed in the training population) captured by the trained models. Generalisation errors evaluated for the models investigated in this study highlight the ability of IG-PO to generalise to unseen shapes better than PCA for three out of the five cardiovascular structures of interest (i.e., LV, LA and RA). 

Furthermore, the clinical relevance of the virtual cohorts synthesised using all the methods investigated in this study was evaluated with respect to the preservation of crucial cardiac volumetric indices relative to the actual population observed from the UK Biobank. The metric used for the same, namely, the clinical acceptance rate (see section~\ref{clinrev}) highlighted the ability of the four models investigated within the proposed generative shape compositional framework (i.e. IG-CO, IG-PO, DG-CO, and DG-PO) to preserve clinically relevant cardiac indices within a $90\%$ interval of the values observed in the natural population, in a higher proportion of their virtual cohort samples than provided by PCA. This helps to further demonstrate the ability of the proposed shape compositional framework to generate better quality cardiac virtual cohorts than PCA, even in the presence of \textit{ missing/partially overlapped} training data.

PCA-based statistical shape modelling is a powerful tool for synthesising virtual populations of anatomical structures~\cite{young2009computational} and has been extensively explored for this purpose in several previous studies \cite{rodero2021linking, romero2021clinically}. However, a fundamental limitation of PCA-based SSMs' inability is to learn latent/low-dimensional shape representations from \textit{missing/partially overlapping} training data. Thus, such models require all parts/structures to be available for samples in the training population to synthesise complete multipart shape assemblies (e.g. whole heart shapes considered in this study). Additionally, PCA-based models are linear projections of shape data onto a lower-dimensional sub-space and cannot effectively capture nonlinear variations in shapes. This results in limited generalisation capacity and specificity for statistical shape models trained using such approaches and, correspondingly, limits the anatomical plausibility of synthesised virtual cohorts. gcVAE-based generative shape models, on the other hand, can capture nonlinear variations in shapes, yielding virtual mates with higher specificity/anatomical plausibility (see Tables~\ref{tab:specificity} and~\ref{tab:clinical-accept}) and generalising better to unseen shapes (refer to Table~\ref{tab:generalization_error}). 

Among the gcVAE-based shape models investigated in this study, the dependent generator models (DG-CO and DG-PO) are based on a multichannel VAE graph convolutional \cite{antelmi2019sparse}, which can capture nonlinear correlations between all cardiovascular structures of interest, by learning a shared latent space across all systems. This enables the dependent generator models to generate virtual cohorts with better anatomical specificity/plausibility than the independent generator models (see Table~\ref{tab: specificity}). The higher specificity of dependent generator models is further complemented by their ability to preserve key cardiac clinical indices in the virtual cohorts synthesised than PCA and, in some cases, their independent generator counterparts, as evidenced by the higher clinical acceptance rates achieved for each cardiac index evaluated (refer to Table~\ref{tab:clinical-accept}). Improved specificity, however, comes at the cost of generalisability and the overall variability in shape captured by the dependent generator models as the learnt latent space is constrained to explain the shared variation of all structures in the whole heart shape assemblies observed across the training population. 

Conversely, the independent generator models do not capture correlations between structures as they use independent part-wise VAEs. However, their dedicated part-wise latent spaces provide greater flexibility than their dependent generator counterparts, enabling them to capture greater variability in shape for the structures of interest. This is reflected in the lower generalisation errors achieved by the independent generator models (see Table~\ref{tab:generalization_error}) and the wider range of values observed for the relevant cardiac indices in the virtual cohorts synthesised (see Figure~\ref{fig:cardiac-indexes}), relative to their dependent generator counterparts. 

In this study, we incorporate both generators into our compositional pipeline instead of using either one to give our model more suitability/applicability for responding to the different scenarios in conducting ISTs or in silico simulation studies.
The dependent generator model and cardiac \textit{virtual chimaera} cohorts synthesised thereof are better suited to ISTs where greater statistical fidelity is required in the enrolled virtual patients/\textit{chimaeras} in terms of key/relevant anatomical phenotypes. 
The virtual cohorts synthesised by this generator may be used in public health planning, personalised treatment, and medical device evaluation in specific target patient groups.
Conversely, the independent generator model is better suited to exploratory ISTs, where the aim is to investigate 'what if' scenarios and assess the performance of medical devices in niche sub-populations that express phenotypic traits that are concentrated in the tails of the general population distribution, i.e. in other words, where, virtual cohorts with greater anatomical variability are desired. 
In summary, the trade-off between specificity and diversity required in a virtual population varies depending on the specific research objectives and its context of use. The construction of a well-rounded pipeline can lead to solid applicability in clinical practice.

Previous studies on generative shape composition learning~\cite{li2020learning,li2021editvae} within the computer vision domain have relied on fully supervised learning to spatially compose synthesised parts into multipart shape assemblies by estimating rigid or affine transformations, as stated previously. Such approaches work well when the shape assemblies of interest are rigid bodies/structures (e.g. chairs, tables, etc.). However, the anatomical structures of soft tissues are deformable and exhibit significant non-linear variations in shape between individuals in a population. Consequently, spatially composing multiple such structures into anatomically plausible shape assemblies requires affine and nonrigid registration. In other words, generative shape composition learning for multipart/multiorgan anatomies requires the synthesised structures to be treated as a recomposable set of deformable parts. We address this challenge by introducing a self-supervised affine and nonrigid spatial composition network. This alleviates the need for ground-truth transformations and composed multipart shape assemblies to be available \textit{a priori}. As shown in Table~\ref{tab:share_vertex}, the non-rigid registration component of the proposed spatial composition network yields significant improvements over its purely affine counterpart in terms of coherence achieved at shared boundaries between adjacent structures in the assemblies composed of the whole heart shape.
Furthermore, the self-supervised learning approach used to train the spatial composition network is driven by weak labels defined by the shared vertices between all adjacent structures with shared boundaries. Hence, the composition network can be trained with parts/structures that are synthesised independently of each other. This is central to enabling the proposed generative shape composition framework to be trained with \textit{ partly overlapping} data (ie, where the training population comprises patient data with missing parts/structures).

\section{Limitations, Challenges and Future Work}
Although the proposed approach facilitates the synthesis of virtual heart cohorts using missing / partially overlapping training data, some limitations, challenges and potential for future improvements remain. These include -  
\subsubsection{Diverse Topology} The current generative shape compositional framework requires all input shapes, represented as surface meshes, to share point-wise correspondence and comprise the same graph topology (i.e., mesh triangle connectivity should be identical). This requires coregistration of all meshes of the corresponding parts for all patients included in the training, validation, and test populations before training or evaluating any of the components of the proposed framework. This co-registration preprocessing step helps establish point-wise spatial correspondence and maintain a fixed graph topology across all samples, but also limits the utility of the proposed generative shape compositional approach to those applications where anatomical shape correspondence exists and can be estimated. This precludes the application of the present framework from modelling organ shapes where pathology-driven topological changes are present and diverse across patient populations (as there is no notion of anatomical correspondence in such a scenario). Expanding the current approach to accommodate variable topology in anatomical structures between patients/input samples and synthesising anatomically plausible virtual populations would broaden the range of applications in computational medicine and \textit{in silico} trials for which the approach would be suitable. 
\subsubsection{Topological Guarantee} The current compositional framework of generative shapes does not guarantee that the topology of the virtual chimaeras synthesised is preserved relative to the native anatomy, as no explicit constraint enforces the same. Although the spatial composition network achieves low errors following the alignment/composition of individual cardiac structures into a whole-heart shape assembly, minor topological defects, such as localised intersections/holes between adjacent frames, remain in some virtual chimaera instances. In the context of conducting \textit{in silico} trials using cardiac virtual chimaera cohorts synthesised by the proposed approach, these topological errors must be fixed using appropriate mesh/geometry processing techniques (which is feasible) before computational volumetric meshes that are usable in biomechanistic simulations can be created from the same. Such issues can be addressed by introducing additional geometric and/or topological constraints into the learning process, which we will explore in future work. 
\subsubsection{Conditional Generation} The current work focusses on the unconditional generation of virtual heart cohorts. This may limit its application in ISTs / in silico studies that require specific characteristics of the target population. Typically, an essential aspect of patient recruitment in actual clinical trials used to assess device performance and generate regulatory evidence for device approval is the precise definition of the inclusion and exclusion criteria for the problem. These criteria define the target patient population deemed appropriate/safe to evaluate the performance of the device of interest. Therefore, an approach that facilitates the controllable synthesis of virtual anatomies is desirable in many scenarios of IST. In the future, we will extend our framework to a conditional-generative model to meet this requirement.
\subsubsection{Real Clinical Scenario} Finally, the current study only emulates the learning scenario to synthesise whole-heart shape assemblies using missing/partially overlapping training data from different patient populations (i.e., only UK Biobank data are used throughout this study). Future work will explore the combination of anatomical structures extracted from multimodal imaging data acquired between different patient populations (e.g., CTA aortic vessel available from a clinical trial and cine MRI left ventricle available in UK Biobank) to synthesise virtual heart chimaera cohorts.

\section{Conclusion}
A generative shape modelling framework is proposed to build virtual cardiac populations. A vital contribution of the study is to treat the synthesis of multipart objects, such as whole-heart shape assemblies, as one of the learning processes for generative shape composition. The proposed approach can synthesise complete groups of whole heart shapes, using \textit{partially overlapping} training data, where all cardiovascular structures of interest are unavailable for all patients that comprise the training population. This demonstrates its potential to combine \textit{partially overlapping} anatomical structures from disparate databases and patient populations to synthesise plausible virtual heart cohorts. We explore two generative shape modelling schemes within the proposed framework: the independent and dependent generators. The former facilitates the generation of more diverse virtual heart cohorts regarding variability in the shapes of cardiovascular structures and their corresponding clinically relevant volumetric indices. The latter provides greater statistical fidelity in specificity, resulting in more anatomically plausible cardiac virtual cohorts. Although the synthesis of virtual heart cohorts is the focus of this proof-of-concept study, the proposed generative shape compositional framework is generic. It may be employed to synthesise virtual affiliates of other multipart organs or multiorgan ensembles (e.g., lungs and their associated airways, abdominal organs, the complete spine, etc.). This study is an essential step toward integrating anatomical shape information from disparate, multimodal data sets and diverse patient populations to synthesise \textit{ virtual heart chimaera} cohorts suitable for conducting \textit{in silico} trials of medical devices.

\bibliographystyle{IEEEtran}
\bibliography{references.bib}

\end{document}